\documentclass[twocolumn,english,reprint,amsmath,amssymb,aps,groupaddress,NOlinenumbers,showpacs,prl,superscriotaddress]{revtex4-1}

\usepackage{float}
\usepackage{graphicx}
\usepackage{dcolumn}
\usepackage{bm}

\usepackage{color}
\usepackage{multirow}
\usepackage{enumerate}

\begin{document}

\title{Mesoscopic limit cycles in coupled nanolasers}

\author{Mathias Marconi}
\altaffiliation{Current address: Institut de Physique de Nice, Universit\'e C\^ote d’Azur, CNRS, Nice, France}
\affiliation{Centre de Nanosciences et de Nanotechnologies, CNRS, Universit\'e Paris-Sud, Universit\'e Paris-Saclay, 10 Boulevard Thomas Gobert, 91120 Palaiseau, France}

\author{Fabrice Raineri}
\altaffiliation{Universit\'e de Paris, Univ Paris Diderot, 75013 Paris, France}
\affiliation{Centre de Nanosciences et de Nanotechnologies, CNRS, Universit\'e Paris-Sud, Universit\'e Paris-Saclay, 10 Boulevard Thomas Gobert, 91120 Palaiseau, France}

\author{Ariel Levenson}
\affiliation{Centre de Nanosciences et de Nanotechnologies, CNRS, Universit\'e Paris-Sud, Universit\'e Paris-Saclay, 10 Boulevard Thomas Gobert, 91120 Palaiseau, France}

\author{Alejandro M. Yacomotti}
\email{alejandro.giacomotti@c2n.upsaclay.fr}
\affiliation{Centre de Nanosciences et de Nanotechnologies, CNRS, Universit\'e Paris-Sud, Universit\'e Paris-Saclay, 10 Boulevard Thomas Gobert, 91120 Palaiseau, France}

\author{Julien Javaloyes}
\affiliation{Departament de F\'isica, Universitat de les illes Balears, C/ Valdemossa km 7.5, 07122 Mallorca, Spain }


\author{Si H. Pan}
\affiliation{Department of Electrical and Computer Engineering, University of California, 9500 Gilman Drive, La Jolla, San Diego, California 92093, USA}
\author{Abdelkrim El Amili}
\affiliation{Department of Electrical and Computer Engineering, University of California, 9500 Gilman Drive, La Jolla, San Diego, California 92093, USA}
\author{Yeshaiahu Fainman}
\affiliation{Department of Electrical and Computer Engineering, University of California, 9500 Gilman Drive, La Jolla, San Diego, California 92093, USA}

\date{\today}


\begin{abstract}

Two coupled semiconductor nanolasers exhibit a mode switching transition, theoretically characterized by limit cycle --or mode-beating-- oscillations. Their decay rate is vanishingly small in the thermodynamic limit, i.e. when the spontaneous emission noise $\beta$-factor tends to zero. 
We provide experimental evidence of mesoscopic limit cycles --with $\sim 10^3$  intracavity photons-- through photon statistics measurements. We first show that the order parameter quantifying the limit cycle amplitude can be reconstructed from the mode intensity statistics. 
As a main result we observe a maximum of the averaged amplitude at the mode switching, accounting for limit cycle oscillations. We finally relate 
this maximum to a dip of mode cross-correlations, reaching a minimum of $g_{ij}^{(2)}=2/3$, which we show to be a mesoscopic limit. 
Coupled nanolasers are thus an appealing testbed for the investigation of spontaneous breaking of time-translation symmetry in presence of strong quantum fluctuations.

\end{abstract}
\maketitle

How quantum fluctuations affect nonequilibrium periodic orbits? 
This question, intimately related to the spontaneous breaking of time translation symmetry, has strongly motivated a large community of physicists in the last few years. 
Although the spontaneous time symmetry breaking is well known in classical nonlinear dynamical science \cite{Guckenheimer:1986aa}, its realization in the quantum world has been a subject of debate. In a seminal paper \cite{PhysRevLett.109.160401}, F. Wilczek pointed out the existence of quantum periodic motion in a time-invariant Hamiltonian, launching a new field of research known as time crystals: the time counterparts of spatial crystals, for which the continuous spatial translation symmetry is spontaneously broken. Since then many efforts have been devoted to understand and implement time crystals in different domains such as condensed matter and QED systems (see, e.g., Ref. \cite{Sacha_2017} for a review).

\begin{figure*}[ht!]
\begin{centering}
\includegraphics[trim=2cm 5cm 1cm 1cm,clip=true,scale=0.6,angle=0,origin=c]{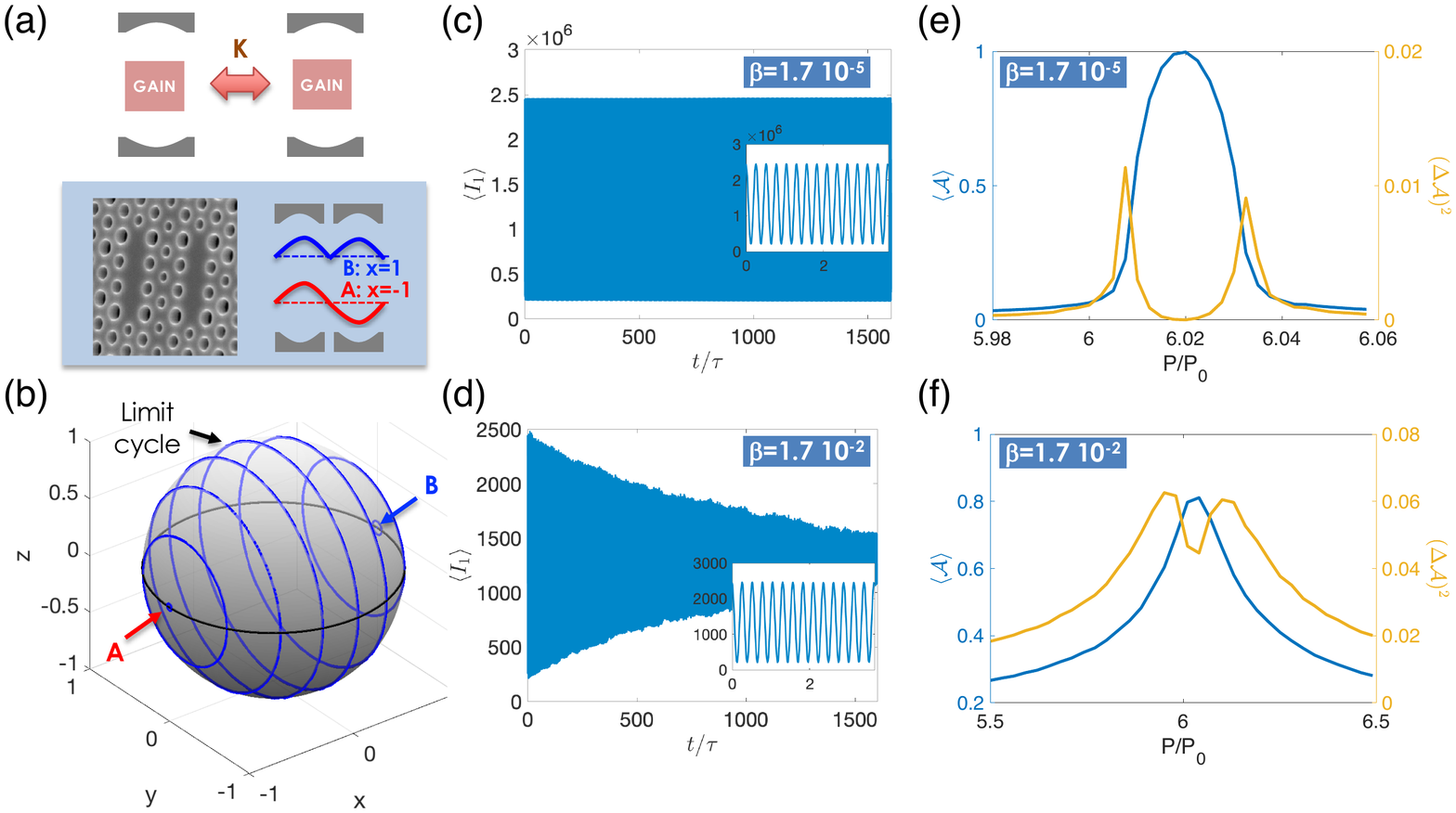}
\par\end{centering}
\centering{}\caption{Mode beating limit cycles in two strongly coupled nanolasers subjected to spontaneous emission noise. (a) Schematics of two evanescently coupled laser nanocavities, with coupling strength $K$. Bottom frame: SEM image of the utilized InP photonic crystal coupled lasers; right: the two modes of the photonic molecule (B: bonding, A: antibonding) are split in energy. (b) Bloch sphere, showing noiseless limit cycles close to a switching point ($F_{a_{1,2}}(t)=0$ in Eqs. \ref{eq:aLR}) ; blue trajectories correspond to orbits increasing pumping power from $P/P_0=6.008$ (fixed point B: $x=1$), $6.010, 6.012, 6.016$ (limit cycles, $x>0$), $6.020$ (perfect mode beating limit cycle: $x\approx 0$) , $6.024, 6.028, 6.032$ (limit cycles, $x<0$), and $6.036$ (fixed point A: $x=-1$); other parameters (Eqs. \ref{eq:aLR}-\ref{eq:nLR}) are $K=12\kappa$, $\gamma=0.05 \kappa$, $\alpha=7$, $\gamma_\parallel=2.2$ GHz, $\gamma_{tot}=5$ GHz, $\kappa=140.84$ GHz, $n_0=10^{18} \text{cm}^{-3}\times V_a$ with $V_a=0.016\times 10^{-12} \text{cm}^{3}$. (c)-(d) Ensemble average of the intensity in the cavity 1 for $P/P_0=6.015$; the average is taken over 100 different noise realizations and the same initial conditions (see text). Insets: zoom over few oscillation periods. (e)-(f) Mean value and fluctuations of the order parameter as a function of the pump power $P$ normalized to the transparency pump $P_0$. (c) and (e) $\beta=1.7\times 10^{-5}$ and $V_a=16\times 10^{-12} \text{cm}^{-3}$. (d) and (f) $\beta=0.017$ and $V_a=0.016\times 10^{-12} \text{cm}^{3}$. Other parameters of the Langevin terms are $F_P=1.03$ and $B=3\times 10^{10} \text{cm}^3s^{-1}$. 
\label{Fig_switching}}
\end{figure*}

Recently, F. Iemini et al. have proposed a new class of dissipative time crystals called boundary time crystals (BTC's) \cite{PhysRevLett.121.035301}. In contrast to Floquet time crystals --or $\pi$-spin glasses-- \cite{PhysRevA.91.033617,PhysRevLett.117.090402,PhysRevB.96.115127} subjected to a periodic forcing, 
in BTC's the Hamiltonian is time-independent, and it is the \textit{continuous} time symmetry which is spontaneously broken in a small though macroscopic fraction of a many body quantum system. The prediction is that, in the thermodynamic limit, a periodic solution emerges whose decay rate tends to zero,  i.e. the amplitude of the oscillations becomes constant in time. 
Such a divergent time scale is related to a closure of a Liouvillian gap in the thermodynamical limit, hence to a dissipative phase transition \cite{PhysRevA.98.042118}.  In a BTC this decay rate is associated to the lowest eigenvalue with nonzero imaginary part \cite{PhysRevLett.121.035301}.
Hence, the persistent oscillations are associated to the spontaneous symmetry breaking since they only take place in the thermodynamic limit. The model developed in Ref.  \cite{PhysRevLett.121.035301} accounts for cooperative emission in cavities, which can be realized by cold atoms in a cavity subjected to Raman driving. In addition, a number of many body limit cycles could be classified as BTCs \cite{PhysRevLett.111.073603,PhysRevLett.116.143603,PhysRevLett.110.163605}. An important conclusion is that the existence of BTCs is to be experimentally tested, since limit cycles might not survive in the presence of fluctuations. 

In this work we propose coupled nanolasers (examples of recent realizations can be found in Refs. \cite{Hamel:2015vn,Deka:17,PhysRevX.8.011013}) as testbeds for limit cycles subjected to strong quantum noise --in this case due to spontaneous emission-- 
and provide experimental evidence on the existence of limit cycles with a thousand photons inside the cavities. 

Lasers are fascinating workbenches to study non-equilibrium statistical mechanics. A paradagmatic 
example, realized since the early days of laser theory, is the second order phase transition at the oscillation threshold in the thermodynamic limit, i.e. for vanishingly small spontaneous emission $\beta$ factor \cite{PhysRevA.2.1170, RevModPhys.47.67,PhysRevA.50.4318}; intracavity photon number scales as 
$\beta^{-1}$, which can be identified as the thermodynamic parameter \cite{PhysRevA.50.4318}. Here we explore limit cycle oscillations that emerge as mode beating when the two eigenmodes of the two coupled nanolasers operate simultaneously. Specifically, this occurs at a mode switching transition between the bonding and anti-bonding modes of a photonic crystal molecule \cite{Marconi:16}.

We consider the photon statistics around a switching transition from the bonding ($B$) to the anti-bonding ($A$) modes of a nanolaser dimer formed by two evanescently coupled semiconductor nanocavities (coupling constant $K$, see Fig. \ref{Fig_switching}a) \cite{Marconi:16,Hamel:2015vn} as the pump is increased. At the switching point, the high energy (blue-shifted, here $B$) mode switches off, and simultaneously the fundamental (red-shifted, here $A$) mode switches on \cite{Marconi:16}. 

Theoretically, lasers can be described by a quantum master equation using the density matrix approach \cite{scully_zubairy_1997,takemura2019low}. Much more simplified models have been used in the past: among them, the semiclassical laser theory --which neglects quantum fluctuations-- has the status of a mean field model in statistical mechanics \cite{PhysRevA.50.4318}, which is a rough approximation for semiconductor nanolasers. A more realistic description needs to incorporate spontaneous emission fluctuations produced by the semiconductor emitters (such as quantum wells, QWs), which can be added to the semiclassical model in the form of Langevin noise terms. Two coupled nanolasers containing QWs can be thus modeled by the following nonlinear coupled stochastic differential equations \cite{Hamel:2015vn,Marconi:16,PhysRevX.8.011013}: 
\begin{align}
\dot{a}_{1,2}  & = \left( \frac{1+i\alpha}{2} G_{1,2} -\kappa \right) a_{1,2}+ \left(\gamma+iK\right)a_{2,1} +F_{a_{1,2}}(t) \label{eq:aLR}\\
\dot{n}_{1,2}  &=  P -\gamma_{tot} n_{1,2}- G_{1,2}|a_{1,2}|^2 \label{eq:nLR}
\end{align}
where $|a_{i,j}|^2=I_{i,j}$ and $n$ are normalized as the photon and carrier numbers in the cavities, respectively, $\kappa$ is the cavity loss rate, $\alpha$ the Henry factor, $P$ the pump rate and $\gamma_{tot}$ is the total carrier recombination rate. The complex inter-cavity coupling constant quantifies frequency ($K$) and loss ($\gamma$) splitting as a result of the evanescent coupling. $G_{1,2}$ = $\gamma_{\parallel}\beta (n_{1,2}-n_0)$ is the gain, $\gamma_{\parallel}$ is the two-level radiative recombination rate and $n_0$ the carrier number at transparency. $F_{a_i}(t)$ are Langevin noise terms accounting for spontaneous emission with rate  $R_{sp}= \beta F_P B n_{1,2}^2/V_a$ where $B$ is the bimolecular radiative recombination rate, $F_P$ the Purcell factor and $V_a$ the volume of the active medium.  We make the common assumption of uncorrelated (white) noise, i.e. $\langle F_\mu(t) F_\nu(t')\rangle=2D_{\mu \nu}\delta(t-t')$, where $D_{\mu \nu}$ are the following diffusion coefficients: $2D_{a_ia_i}=2D_{a_i^*a_i^*}=0$, $2D_{a_ia_i^*}=2D_{a_i^*a_i}=R_{sp}$, and zero otherwise. Importantly, the spontaneous emission factor  $\beta$ is related to the intracavity saturation photon number as $I_{sat}=\gamma_{tot}/\gamma_{\parallel}\beta$; $\beta^{-1}$ can thus be identified as the thermodynamic parameter, since the characteristic photon number scales as $\beta^{-1}$. 

\begin{figure}[ht!]
\begin{centering}
\includegraphics[trim=1cm 1cm 0cm 1cm,scale=0.47]{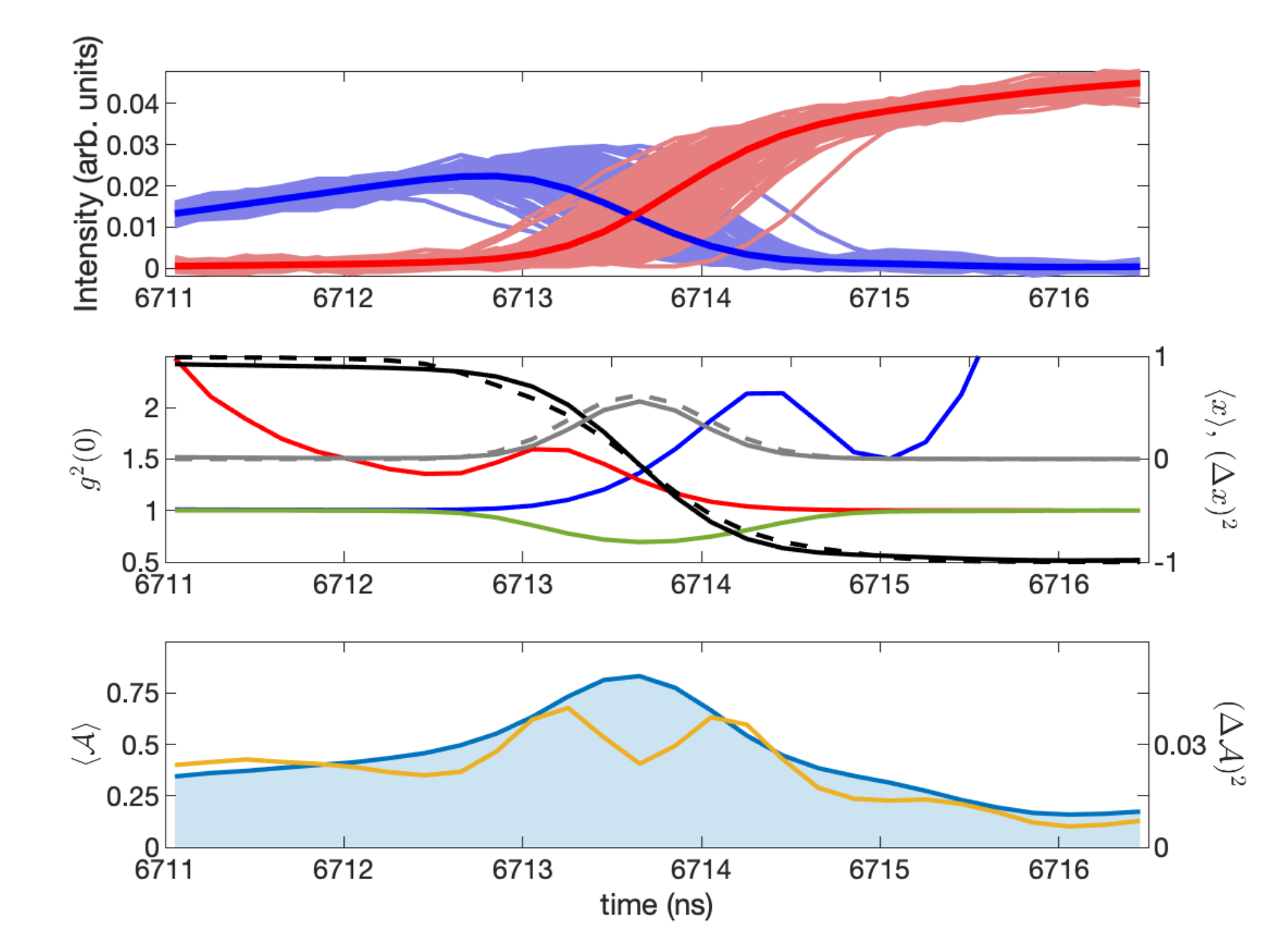}
\par\end{centering}
\centering{}\caption{Experimental time traces as the pump power is quasi-statically ramped up in time. Top panel: Intensity traces for bonding (blue) and antibonding (red) modes measured with 600 MHz-APD detectors. Pump ramp duration=6 ns. Thick lines: average corresponding to $10^4$ time traces. Middle panel: second order correlations (left axis) and the two lowest moments of the mode population imbalance $x$ (right axis). Blue: $g^{(2)}_{BB}$, red: $g^{(2)}_{AA}$ and green: $g^{(2)}_{BA}$. Black: mean value, and grey: variance of $x$. Solid lines show the results using the full statistics; dashed lines compute the moments from $g^{(2)}_{ij}$ (Eqs. S3-S5 of the Supplementary Material), thus neglecting correlations between $I$ and $x$. Bottom panel: two first moments of the order parameter $\mathcal{A}$. Blue: mean value, $\langle \mathcal{A}\rangle$; yellow: variance, $(\Delta \mathcal{A})^2$. }
 \label{Fig_g2}
\end{figure}
The two linear eigenmodes of Eqs.~\ref{eq:aLR} are $a_B=(a_1+a_2)/\sqrt{2}$ and $a_{A}=(a_1-a_2)/\sqrt{2}$, corresponding to bonding and anti-bonding modes of the coupled cavities system, respectively (Fig. \ref{Fig_switching}a). As it has been shown elsewhere \cite{PhysRevX.8.011013}, the dynamics of this system can be separated in two subset of variables: the total intensity and carrier number on one side, and the relative intensities and phases of the cavities on the other side, which can be recast on the Bloch sphere as $\theta=2\arctan\left(\sqrt{I_2/I_1} \right)\in\left[0,\pi\right]$ and $\Phi=\psi_{1}-\psi_{2}$, where $a_j=\sqrt{I_{j}}\exp\left(i\psi_{j}\right)$. Remarkably, the $x$-coordinate of the Bloch sphere is nothing but the mode population imbalance, $x=(I_B-I_{A})/(I_B+I_{A})$, where $I_B$ and $I_{A}$ are the intensities of the two eigenmodes. 

Above laser threshold, the laser molecule operates in the mode with higher net gain, which in our case has been designed to be the $B$-mode. Experimentally, a switching transition is observed, where the $B$-mode switches off, and the $A$-mode switches on as the pump power is ramped up \cite{Marconi:16}. Indeed, Eqs. \ref{eq:aLR}-\ref{eq:nLR} show mode switching dynamics. Interestingly, the mode switching transition is mediated by the emergence of a limit cycle in the thermodynamic limit, $\beta^{-1}\rightarrow \infty$ \cite{Marconi:16}: the $B$-mode loses stability expelling a limit cycle at a first Hopf bifurcation ($x= 1$, Fig. \ref{Fig_switching}b); these oscillations account for mode beating. The limit cycle amplitude rapidly increases up to a perfect mode beating situation in which both modes have the same intensity (dual-frequency laser), and each cavity intensity experiences $100\%$-contrast oscillation ($x= 0$, Fig. \ref{Fig_switching}b). Further increasing the pump parameter the limit cycle shrinks and coalesces at a second Hopf bifurcation, leading to a stable fixed point on the Bloch sphere corresponding to the $A$-mode ($x= -1$, Fig. \ref{Fig_switching}b). 

The limit cycle oscillations are long-lasting solutions of the mean field limit: the amplitude decay rate tends to zero. In the presence of noise, fluctuations increase the decay rate. We have quantified such an effect through simulations of the Langevin equations with different $\beta$-factors accounting for different spontaneous emission rates. In Fig.  \ref{Fig_switching}c-d we show the ensemble average of the cavity 1-intensity, $\langle I_1\rangle$ for $\beta=1.7\times10^{-5}$ (Fig. \ref{Fig_switching}c) and $\beta=1.7\times10^{-2}$ (Fig. \ref{Fig_switching}d). In these simulations the initial conditions correspond to a maximum of photon number in cavity 1, i.e. $\Phi=0$ or $\pi$, and $\theta\leq \pi/2$. Figure \ref{Fig_switching}c corresponds to a macroscopic laser cavity: the effect of noise on the limit cycle is small, and the amplitude does not decay in the whole time window used for the calculations. However,  in  Fig. \ref{Fig_switching}d we observe a drastic reduction in the decay time of the amplitude; still, the limit cycle undergos thousands of oscillations during the damping time. It is important to point out that the period of oscillations is $T=\pi/K=0.26$ in units of the cavity photon lifetime, corresponding to a frequency of $f=545$ GHz in our example of strongly coupled cavities. Such a high frequency combined with a low output photon number, rule out the possibility of the direct observation of the limit cycle. However, we will show below that the order parameter accounting for the limit cycle formation can be quantified through the mode intensity statistics. 

By construction, the limit cycle amplitude on the Bloch sphere of Fig. \ref{Fig_switching}b is 
\begin{equation}
\mathcal{A}  = \sqrt{1-x^2}
\label{eq:A}
\end{equation}
which is the natural order parameter for the limit cycle. Equation  \ref{eq:A} simply states that the limit cycle vanishes for single mode operation, $x=\pm 1$, and reaches a maximum order of $\mathcal{A}=1$ for the \lq \lq meridian" limit cycle ($\Psi=\pi/2$) in the thermodynamic limit. 
In Fig. \ref{Fig_switching}e-f we show the mean value and variance of the order parameter as a function of the pump. Clearly,  $\langle \mathcal{A}\rangle $ reaches a maximum at $P/P_0\approx 6.02$ for $\beta=1.7\times10^{-5}$ (Fig. \ref{Fig_switching}e) with two fluctuation maxima at the bifurcation points. In the \lq \lq nanolaser" case, $\beta=1.7\times 10^{-2}$, $\langle \mathcal{A}\rangle $ still presents a maximum at the mode switching point, but its value is smaller, $\langle \mathcal{A}\rangle \approx 0.8$, and the pumping range for nonzero $\langle \mathcal{A}\rangle $ is broadened with respect to the \lq \lq macroscopic" case. This last point is important since
the nanolaser regime enhances the pumping interval of existence of the limit cycle. Note that the statistics of $\mathcal{A}$ can be obtained through the mode-intensity statistics. In the following we will present our experimental results showing the increase of the limit cycle amplitude at the mode switching point, together with an increase of the amplitude fluctuations at both sides, which is the signature of a limit cycle bifurcation. 



\begin{figure}[ht!]
\begin{centering}
\includegraphics[trim=8cm 0cm 0cm 0cm,clip=true,scale=0.45,angle=0,origin=c]{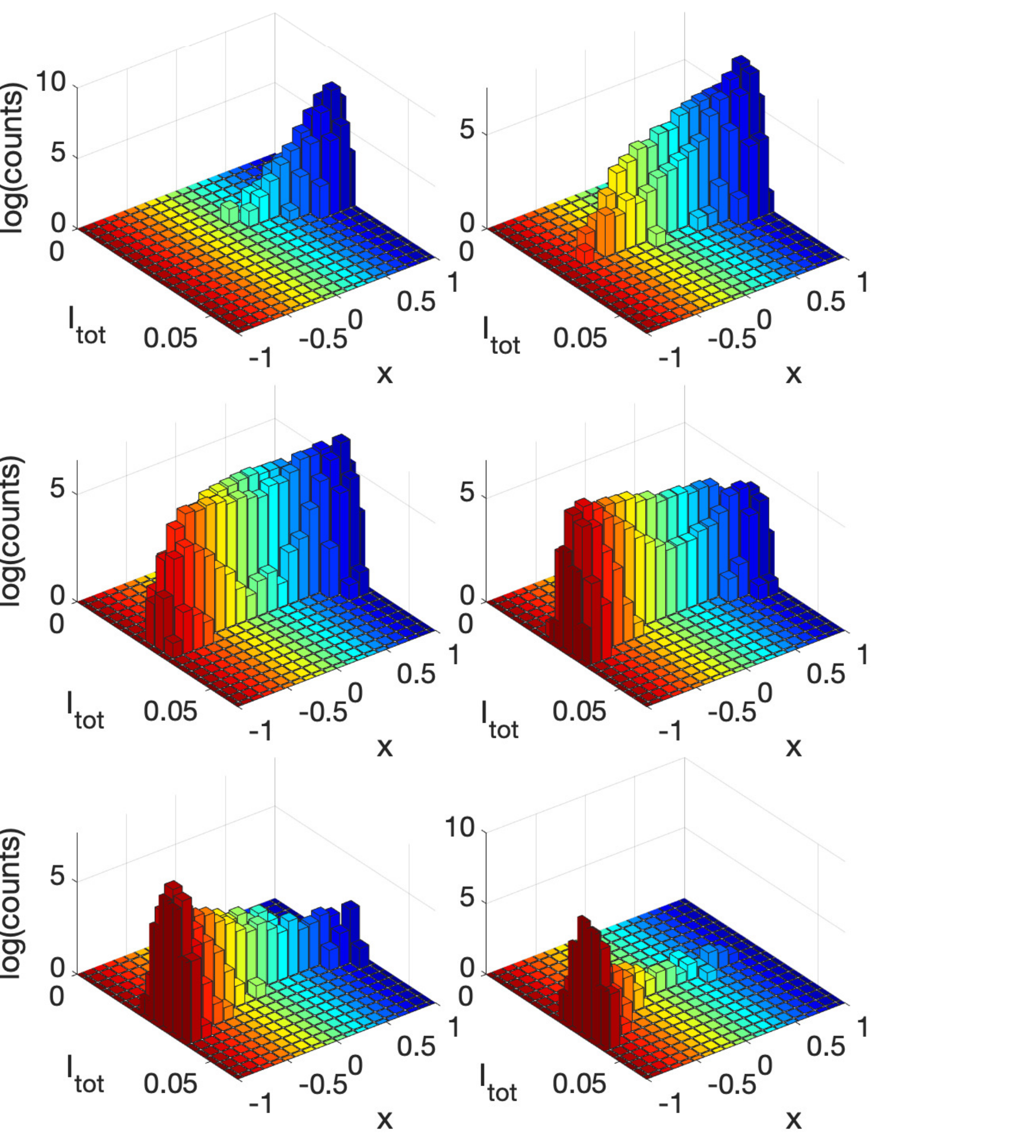}
\par\end{centering}
\centering{}\caption{Joint histograms $P(x,I_{tot}; t)$ corresponding to the rumping-up intensity traces of Fig. \ref{Fig_g2}. Frames correspond to increasing pump powers at different times: from bottom-right to top-left corners t=6712.7, 6713.1, 6713.5, 6713.9, 6714.3, 6714.7. It can be observed that nearly flat $x$-statistics occur at the switching point, i.e. t=6713.5-6713.9. $x$-distributions are exponentially decaying from $x=1$ to $x=-1$ for small pump powers (t=6712.7, top-left), and from $x=-1$ to $x=1$ for large pump powers (t=6714.7, bottom-right). 
 \label{Fig_hist}}
\end{figure}
Figure  \ref{Fig_g2}a shows the experimental time traces of the two eigenmodes. Modes are detected in the far field, in such a way that their emission can be spatially separated. Mode-intensities are then simultaneously measured using two fast (600 MHz-bandwidth) APD photodetectors as the pump power is ramped up (ramp duration= 6 ns). The time series have been used to reconstruct the statistics of the mode population imbalance (Fig. \ref{Fig_g2}b, right axis). It can be observed that $\langle x \rangle$ has a step-like variation with a zero-crossing --that we refer to as the switching point, $P_s$-- as the pump power is increased. 
The full statistics of $x$ can be used to compute the statistics of $\mathcal{A}$. In Fig. \ref{Fig_g2}c we show the mean value $\langle \mathcal{A}\rangle $ together with the variance $(\Delta \mathcal{A})^2$. We observe a  maximum of $\langle \mathcal{A}\rangle \approx 0.83$ at the switching point, in good agreement with the predictions of the Langevin-semiclassical model (Fig. \ref{Fig_switching}f). In addition, there is a peak of $(\Delta \mathcal{A})^2$ at each side of the switching point, also in agreement with the model. Therefore, our measurements reveal the emergence of a limit cycle, even though the direct measurement of the time oscillations of one cavity-intensity cannot be done due to both the extremely high oscillation frequency (of the order of hundreds of GHz) together with the weak output signals (in the sub-$\mu$W range). 

We point out that the intensity fluctuation dynamics becomes extremely slow at the switching point. As a result the fluctuations could be accurately measured with our APD detectors. Indeed, the time-width of autocorrelation functions is typically $2-3$ ns at the switching point, 
 while the timescales of the system are $\sim 10$ ps for the photons, and $200$ ps for the charge carriers in the QWs. Such a timescale stretching is seemingly due to the critically slowing down of the dynamics close to bifurcation points, which is also predicted by the model. 
 It turns out that the proximity to the bifurcation point is translated into the real part of an eigenvalue approaching zero, which corresponds to a dramatic increase of the timescale. This can be theoretically confirmed in the limit of very strong coupling, where the dynamics can be reduced to a 1D Fokker-Planck equation. We have computed the spectrum of the Fokker-Planck operator, which shows a first exited eigenvalue that vanishes at the switching point (Sec. V, Supplementary Material). In this $K\gg 1$ limit the double bifurcation structure leading to the limit cycle (Fig.\ref{Fig_switching}b) is degenerated to a single parameter $P=P_s$, and only the stochastic dynamics of $x$ is considered. We expect that the Fokker-Planck operator of the full Langevin-semiclassical model should possess, at each Hopf bifurcation, two eigenvalues with nonzero imaginary parts and their real part tending to zero as $\beta\rightarrow 0$. We relate these features to a gapless Liouvillian spectrum having a nonzero imaginary part of a quantum master equation description, as it has been predicted for a large class of limit cycles such as BTCs \cite{PhysRevLett.121.035301}.
 

The strong fluctuations in the limit cycle amplitude come from the strong, non-gaussian fluctuations of the mode population imbalance $x$. In order to further investigate the nature of such fluctuations, we first point out that the semiclassical model predicts exponential equilibrium distributions for $x$  in the $K\gg 1$ limit \cite{PhysRevX.8.011013} (see Supplementary Material, Sec. V), namely 
\begin{equation}
\rho_{eq}(x;\Lambda)=\mathcal{N} e^{-\Lambda x}, 
\label{equilibrium}
\end{equation}
where $\mathcal{N}=\Lambda/\left(2\sinh\Lambda\right)$; $\Lambda$ can be approximated as a linear function of the pump close enough to the switching point, $\Lambda \sim (P/P_s-1)I/\beta^2$ (see Sec. V, Supp. Material). 
The experimental statistical distributions of $x$ are shown in Fig. \ref{Fig_hist}. Flat distributions can be observed at the switching point, in agreement with the theoretical prediction in the mesoscopic regime (Eq. \ref{equilibrium}), with $\Lambda>0$ for $P<P_s$, $\Lambda=0$ at $P=P_s$, and $\Lambda>0$ for $P>P_s$.

\begin{figure}[ht!]
\begin{centering}
\includegraphics[trim=2.5cm 1cm 0cm -0.5cm,clip=true,scale=0.34,angle=0,origin=c]{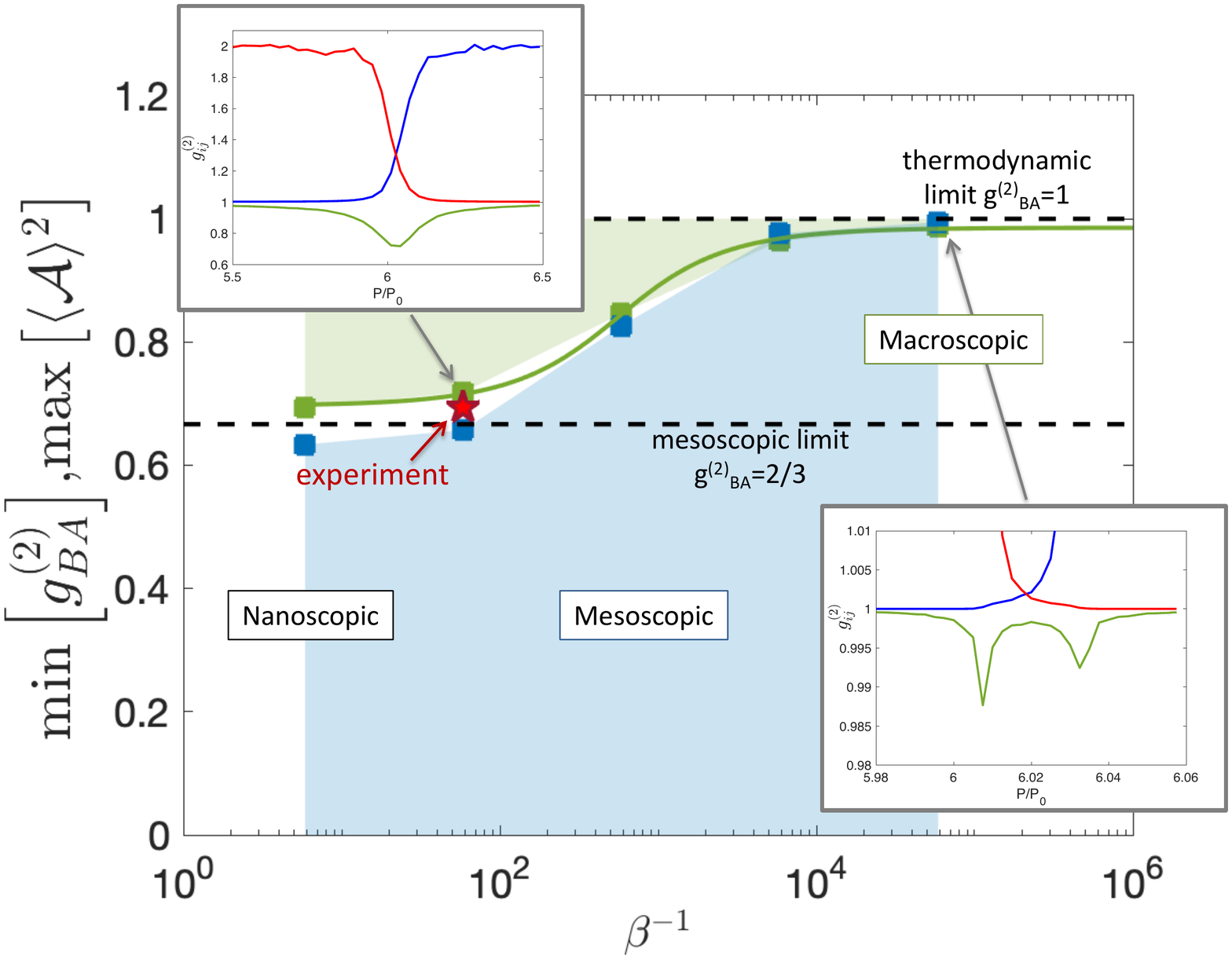}
\par\end{centering}
\centering{}\caption{Minimum of the second order cross correlations (green) and maximum of the squared mean value of the limit cycle amplitude (blue squares) between bonding and anti-bonding modes of a laser dimer for increasing system size $\beta^{-1}$. The thermodynamic and mesoscopic limits are indicated with dashed horizontal lines. Green line: guide to the eye. Inset: zero time delay intensity correlations $g^{(2)}_{ij}$ [$(i,j)=(B,B)$: blue, $(A,A)$: red, and $(B,A)$: green lines] as a function of the pump parameter around the mode switching point, showing: a double dip structure in the macroscopic regime (bottom right), and a single dip in the mesoscopic regime (top left). The experimental result is marked with a red symbol. 
 \label{Fig_vsbeta}}
\end{figure}

Usually, the experimentally accessible quantity is the photon correlations rather than the order parameter. Nevertheless, both of them are related. First we note that the zero time-delay mode cross-correlations reach a limit of $g^{(2)}_{BA}(\tau=0)=2/3$ in the mesoscopic limit cycle regime. This $2/3$ limit can be easily deduced from the relation between the second order coherence and the two lowest order moments of $x$. Under the hypothesis of decorrelated total intensity $I=I_B+I_A$ and $x$-fluctuations, $\langle I x\rangle= \langle I \rangle \langle x\rangle$, it can be shown that 
\begin{equation}
\label{g+-}
g^{(2)}_{BA}=g^{(2)}_{II}\frac{1-\langle x^2\rangle}{1-\langle x\rangle^2}, \\
\end{equation}
where we have removed $\tau=0$ to simplify the notation. We will further assume that the total intensity fluctuations are poissonian, hence $g^{(2)}_{II}=1$, in agreement with our measurements (Fig \ref{Fig_hist}). The ideal case of a flat statistical distribution of $x$ leads to $\langle x \rangle=0$ and $\langle x^2 \rangle=1/3$; hence $g^{(2)}_{BA}=2/3$. This theoretical prediction is also in good agreement with our experimental results: in Fig. \ref{Fig_g2} (middle panel, left axis) we show $g^{(2)}_{AA}$, $g^{(2)}_{BB}$ and $g^{(2)}_{BA}$. Importantly, $g^{(2)}_{BA}\approx 0.7$ is less that unity at the switching point, revealing mode anti-correlations. The mode cross-correlation minimum, min$[g^{(2)}_{BA}]$, is strongly influenced by noise. In Fig. \ref{Fig_vsbeta} we display numerical simulations decreasing the system size from $\beta^{-1}=5.9 \times10^4$ to $5.9$, showing a clear crossover between the macroscopic and the mesoscopic regimes. Interestingly, the cross-correlation functions have a double dip structure in the macroscopic regime, whereas there is a single dip in the mesoscopic regime, in particular for $\beta=0.017$ corresponding to our experimental situation. 

Now, our measurements using classical APD detectors can be compared to photon correlation measurements using single photon detectors, for which the time resolution is as short as $50$ ps for single nanowire single photon detectors (SNSPDs) in the telecommunication band. Cross correlations measurements using SNSPDs under pulse pumping also show a minimum of $g^{(2)}_{BA}\approx 0.7$ close to the switching point (Sec. II, Supplementary Material). 

The cross-correlation is related to the limit cycle amplitude as $\langle \mathcal{A}^2\rangle\approx g^{(2)}_{BA}$ at the switching point (Eq. \ref{g+-}). In addition, neglecting limit cycle amplitude fluctuations we easily get
\begin{equation}
\langle \mathcal{A}\rangle  \approx  \sqrt{g^{(2)}_{BA}}
\label{eq:A_g2}
\end{equation}
close to the switching point; both are shown in Fig. \ref{Fig_vsbeta} for comparison. We point out that $g^{(2)}_{BA}=1$ corresponds to the uncorrelated limit, hence the non-trivial statistical information is contained in the depth of the cross-correlation dip (green area in Fig. \ref{Fig_vsbeta}). Since $\min [g^{(2)}_{BA} ]$ approaches unity in the thermodynamic limit (see Fig. \ref{Fig_vsbeta}, bottom-right panel), no significant statistical information can be extracted from these measurements in the macroscopic regime. In contrast,  the cross-correlation function is a good statistical indicator in the mesoscopic regime,  where $g_{AB}^{(2)}<1$. However, it is important to point out that small modal cross-correlations  ($g_{ij}^{(2)}<1$ with $i\neq j$) is not a sufficient condition for the presence of limit cycles. Indeed, mode anticorrelated fluctuations have already been reported in other photonic systems such as VCSELs or micropillar lasers, but no limit cycle dynamics has been reported in those examples. For instance, polarization switching dynamics in VCSELs have been performed in the past, showing strong mode anticorrelation \cite{PhysRevA.60.4105,PhysRevA.68.033822,PhysRevA.62.033810}, with reported cross correlation functions below $g_{ij}^{(2)}=1/2$ \cite{1397875}. More recently, $g_{ij}^{(2)}<1$ has been shown at the polarization switching of bimodal micro-pillar lasers \cite{PhysRevA.87.053819}, which has been explained as a statistical mixture of a thermal and a coherent state \cite{PhysRevX.7.021045}. 


In conclusion we have shown that mesoscopic limit cycles emerge at the mode switching of a nanolaser dimer. Such limit cycles are mode beating oscillations when two eigenmodes operate simultaneously. This has been possible thanks to photon statistics measurements, which allowed us to compute the order parameter $\mathcal{A}$ which is the amplitude of the limit cycle oscillation. We have shown that a maximum of $\langle \mathcal{A}\rangle$ is observed at the mode switching point, together with two maxima of the order parameter fluctuations at each side of the mode transition, which are signatures of limit cylce bifurcations in the presence of noise, as predicted by a Langevin-semiclassical model. We conjecture that this scenario may support vanishing eigenvalues of the Liouvillian within a quantum master equation description, with a nonzero imaginary part, which has been recently shown to characterize a large family of many body limit cycles \cite{PhysRevLett.121.035301}. In addition, we have related the order parameter to photon correlation measurements, and show that the mesoscopic limit cycle regime is associated with a $2/3$ limit of the mode cross-correlations. Therefore, a coupled nanolaser system proves useful as a testbed for the investigation of limit cycles subjected to strong quantum noise, and the spontaneous breaking of time translation symmetry. 
%
%
%

We acknowledge enlightening discussions with A. Biella, Z. Denis and C. Ciuti. This work has been partially funded by the \lq \lq Investissements d'Avenir" program (Labex NanoSaclay, Grant No. ANR-10-LABX-0035) and the ANR UNIQ DS078.

\providecommand{\noopsort}[1]{}\providecommand{\singleletter}[1]{#1}%

\end{document}